\documentclass[12pt]{article}

\usepackage{amsmath,amsfonts,times,natbib,graphicx}

\newcommand{\boldx} {{\mathbf x}}
\newcommand{\boldu} {{\mathbf u}}

\newcommand{\bolds} {{\mathbf s}}
\newcommand{\boldX} {{\mathbf X}}
\newcommand{\boldZ} {{\mathbf Z}}
\newcommand{\reals} {{\mathbb{R}}}
\newcommand{\sphere} {{\mathbb{S}}}
\newcommand{\boldmu}{\mbox{\boldmath $ \mu$}}
\newcommand{\boldalpha}{\mbox{\boldmath $\alpha$}}

\newcommand{\eqd}{{\stackrel{d}{=}}}
\newcommand{\myfig}[1]{\begin{center} \includegraphics #1 \end{center} }
\newcommand{\Rpkg}[1]{\textit{#1}}

\begin{document}

\author{John P. Nolan\footnote{Supported by contract W911NF-12-1-0385 from the Army Research Office.}, American University }
\title{Models for generalized spherical and \\ related distributions}
\date{30 September 2015}   % early draft at MURI 2013 meeting
\maketitle
\begin{abstract}
A flexible model is developed for multivariate generalized spherical distributions,
i.e. ones with level sets that are star shaped.
To work in dimension above 2 requires tools from computational geometry
and multivariate numerical integration.  In order to simulate
from these star shaped contours, an algorithm to simulate
from general tessellations has been developed that has applications in other situations.
These techniques are implemented in an R package \Rpkg{gensphere}.\footnote{This package
will be made available on CRAN when this paper is published.}
\end{abstract}

% --------------------------------------------------------------------------
\section{Introduction}

There is a need for tractable models for multivariate data with nonstandard
dependence structures.  Our motivation here was to be able to flexibly model distributions
with star-shaped level sets.  An R package \Rpkg{gensphere}
has been developed that allows one to work with these classes of
distributions: specifying flexible shapes for the level sets, computing densities, and
simulating.  A deliberate goal in this process is to have methods and programs that work in dimension
$d \ge 2$, and this requires some methods from computational geometry.  While the original intent focused on star-shaped regions,
some of the tools developed here are useful for other problems, e.g. sampling from more general sets

\cite{fernandez:osiewalski:steel:1995} proposed defining multivariate distributions
from a contour $\mathcal{C}$ (a simple closed curve/surface in $\reals^d$) that is specified
by a contour function $c: \sphere \to [0,\infty)$:
$$\mathcal{C} = \{ c(\bolds) \bolds \, : \, \bolds \in \sphere \}.$$
Here $\sphere =\{ \bolds \in \reals^d : |\boldx|=1 \}$ is the unit sphere, a $(d-1)$-dimensional
surface.
We assume that $c(\bolds)$ is a continuous function.
Figure~\ref{fig:construct2d} shows a 2-dimensional example and Figure~\ref{fig:star3d} shows a
3-dimensional example.

%\begin{figure}
%\myfig{[height=3in]{blob1.pdf}}
%\caption{A contour $\mathcal{C}$ and contour function $c(\bolds)$.}\label{fig:blob1}
%\end{figure}

Let $g:[0,\infty) \to [0,\infty)$ be a nonnegative function and define
\begin{equation} \label{eq:f.def}
f(\boldx) = \begin{cases} g \left( \frac {|\boldx|} {c(\boldx/|\boldx|)} \right) & |\boldx| > 0 \\
                           g(0) & |\boldx|=0
            \end{cases}
\end{equation}
Under integrability conditions discussed below, this will give a probability density function
on $\reals^d$, and the level sets of such a distribution are scalar multiples of $\mathcal{C}$.
Such distributions are also called \textit{homothetic}.
We will call $c(\cdot)$ the \textit{contour function} and $g(\cdot)$ the \textit{radial function}
of the distribution.

Our approach differs a bit from \cite{fernandez:osiewalski:steel:1995} because we take the contour function
$c(\cdot)$  as the basic object, whereas they use  $v(\boldx) := |\boldx|/c(\boldx/|\boldx|)$ as the basic object.
This object is well studied in convex analysis and functional analysis where it is called a \textit{gauge function}
or \textit{Minkowski functional}.
By construction, $v$ is homogeneous:  $v( a \boldx ) = |a| v(\boldx)$.
If $c(\bolds)= 1$, then $\mathcal{C}$ is the unit sphere and $v(\boldx)=|\boldx|$,
so the resulting classes of distributions are the spherical/isotropic distributions.
If $v(\cdot)$ is convex, then $v(\cdot)$ is a norm on $\reals^d$ and $\mathcal{C}$ is
the unit ball in that norm, hence the name $v$-spherical
distributions.   When $v(\cdot)$ is not convex, e.g. an $\ell_p$ ball with $p < 1$, $v(\boldx)$ does not give a norm, so
$\mathcal{C}$ is not strictly speaking a unit ball, but we will still call the resulting distributions
$v$-spherical.

The purpose of this paper is to describe a method of defining a flexible
class of generalized spherical distributions in any dimension $d \ge 2$, and to
describe an R package \Rpkg{gensphere} that implements this method.  The package gives
the ability to
\begin{enumerate}
  \item define a flexible set of contours
  \item carefully tessellate a contour
  \item sample from a tessellation
  \item use a contour and a radial function $g(\cdot)$ to define a generalized spherical distribution
  \item compute the density $f(\cdot)$ given by (\ref{eq:f.def})
  \item approximately simulate from a distribution with density $f(\cdot)$.
\end{enumerate}
The third step above also provides a way to simulate from paths and surfaces unrelated to
generalized spherical laws.

Other references on generalized spherical laws are \cite{arnold:castillo:sarabia:2008},
\cite{kamiya:takemura:kuriki:2008},
\cite{rattihalli:basugade:2009}, \cite{rattihalli:patil:2010},  and \cite{balkema:nolde:2010}.
These papers develop the idea of generalized spherical distributions, but do not provide general purpose software for
working with these distributions and don't cover techniques for working
with higher dimensional models.

% --------------------------------------------------------------------------
\section{Generalized spherical distributions}
%-----------------------------------------------------------------------------
%\subsection{Conditions that guarantee a density function}

For (\ref{eq:f.def}) to be a proper density, it is required that (see \cite{fernandez:osiewalski:steel:1995}, e.g.. (4) and (5))
\begin{equation}\label{eq:norming.constant}
k_{\mathcal{C}}^{-1} := \int_{\sphere} c^d(\bolds) d\bolds \in (0, \infty)
\end{equation}
and
\begin{equation}\label{eq:g.integ}
\int_0^\infty r^{d-1} g(r) dr = k_{\mathcal{C}}.
\end{equation}
We will assume $c(\cdot)$ is continuous on $\sphere$
and that $c(\bolds) \le c_0$.  This guarantees (\ref{eq:norming.constant}) is finite,
though evaluating it may be difficult even when $d=2$, and especially when $d > 2$.

%The polar coordinates version of (\ref{eq:norming.constant}) is
%$ k_{\mathcal{C}}^{-1} := \int_{\Theta} c^d(t(\boldtheta)) |J(\boldtheta)| d\boldtheta \in (0, \infty) $,
%where $\boldtheta = (\theta_1,\ldots,\theta_{d-1}) \in \Theta := (-\pi/2,\pi/2)^{d-2} \times [0,2 \pi)$,
%$t(\boldtheta) = (\sin \theta_1,\linebreak \cos \theta_1 \sin \theta_2,\ldots,(\prod_{j=1}^{d-1} \cos \theta_j) \sin \theta_{d-1})$,
%and $J(\boldtheta) = \prod_{i=1}^{d-2} \cos^{d-1-i} \theta_i$ is the Jacobian of the polar transform.

%  CONSIDER THE APPROACH IN BOAS & ? using vector notation
\subsection{Stochastic representation}
Given any univariate density $h(\cdot)$ on the positive axis, the function
$g(r)=k_{\mathcal{C}} r^{1-d} h(r)$ is a valid radial function. This is the
approach we use in the rest of this paper and in the \Rpkg{gensphere} package.
In this case the positive random variable $R$ with density $h(\cdot)$ gives a stochastic
representation of the generalized spherical random vector:
\begin{equation}\label{eq:stoch.rep}
\boldX \eqd R \boldZ,
\end{equation}
where $\boldZ$ is uniformly distributed on the contour $\mathcal{C}$.
This directly gives a way to simulate $\boldX$ if $\boldZ$ can be simulated.
\cite{balkema:nolde:2010} give a related stochastic representation
$$\boldX \eqd R^* \boldZ^*,$$
where $\boldZ^*$ is uniformly distributed on the unit disk
$\mathcal{D} = \{ \boldx \in \reals^d : v(\boldx) \le 1 \}$.  An advantage of this is
that it is straightforward, though possibly inefficient, to simulate from $\mathcal{D}$
by generating a uniform vector on a rectangle that contains the ball $\mathcal{D}$
and rejecting if $v(\boldx) > 1$.  We prefer to use (\ref{eq:stoch.rep}) and describe
below how to approximately sample from the contour $\mathcal{C}$.

\clearpage
%-----------------------------------------------------------------------------
\subsection{Specification of a contour function}

For modeling purposes, we want a flexible family of functions that can be
used in a variety of problems.
%In particular, we want to be able to model
%star-shaped contours that arise in munitions fragment patterns.
To be able to
include the distributions discussed by the authors cited above, we allow
contour functions of the form
$$c(\bolds) = \sum_{j=1}^{N_1} c_j r_j(\bolds) + \displaystyle{ \frac {1} { \sum_{j=1}^{N_2} c^*_j r^*_j(\bolds) }},$$
where $c_j > 0$, $c^*_j >0$, and $r_j(\cdot)$ and/or $r^*(\cdot)$ are one of the cases discussed below.
\begin{enumerate}
\item $r(\bolds) = 1$, which makes $\mathcal{C}$ the Euclidean ball.

\item $r(\bolds) = c(\bolds|\boldmu,\theta)$ is a cone with peak 1 at center $\boldmu \in \sphere$ and height 0 at the base given
by the circle $\{ \boldx \in \sphere : \boldmu \cdot \boldx = \cos \theta \}$.  It is assumed that $|\theta| \le \pi/2$.

\item $r(\bolds) = c(\bolds|\boldmu,\sigma) = \exp( -t(\bolds)^2/ (2 \sigma^2))$ is a Gaussian bump centered at location
$\boldmu \in \sphere$ and ``standard deviation'' $\sigma>0$. Here $t(\bolds)$ is the distance between $\boldmu$ and
the projection of $\bolds \in \sphere$ linearly onto the plane tangent to $\sphere$ at $\boldmu$.

\item $r^*(\bolds) = \vert \vert \bolds \vert \vert_{\ell^p(\reals^d)}$, $p > 0$.

\item $r^*(\bolds) = \vert \vert A \bolds \vert \vert_{\ell^p(\reals^m)}$, $p > 0$, $A$ an $(m \times d)$ matrix.
This allows a generalized $p$-norm.  If $A$ is $d \times d$ and orthogonal, then the resulting contour
will be a rotation of the standard unit ball in $\ell^p$.  If $A$ is $d \times d$ and not orthogonal,
then the contour will be sheared.  If $m > d$, it will give the $\ell^p$ norm on $\reals^m$
of $A\bolds$.

\item $r^*(\bolds) = (\bolds^T A \bolds)^{1/2}$, where $A$ is a positive definite $(d \times d)$ matrix.  Then
the level curves of the distribution are ellipses.
\end{enumerate}

\begin{figure}
\myfig{[width=\textwidth]{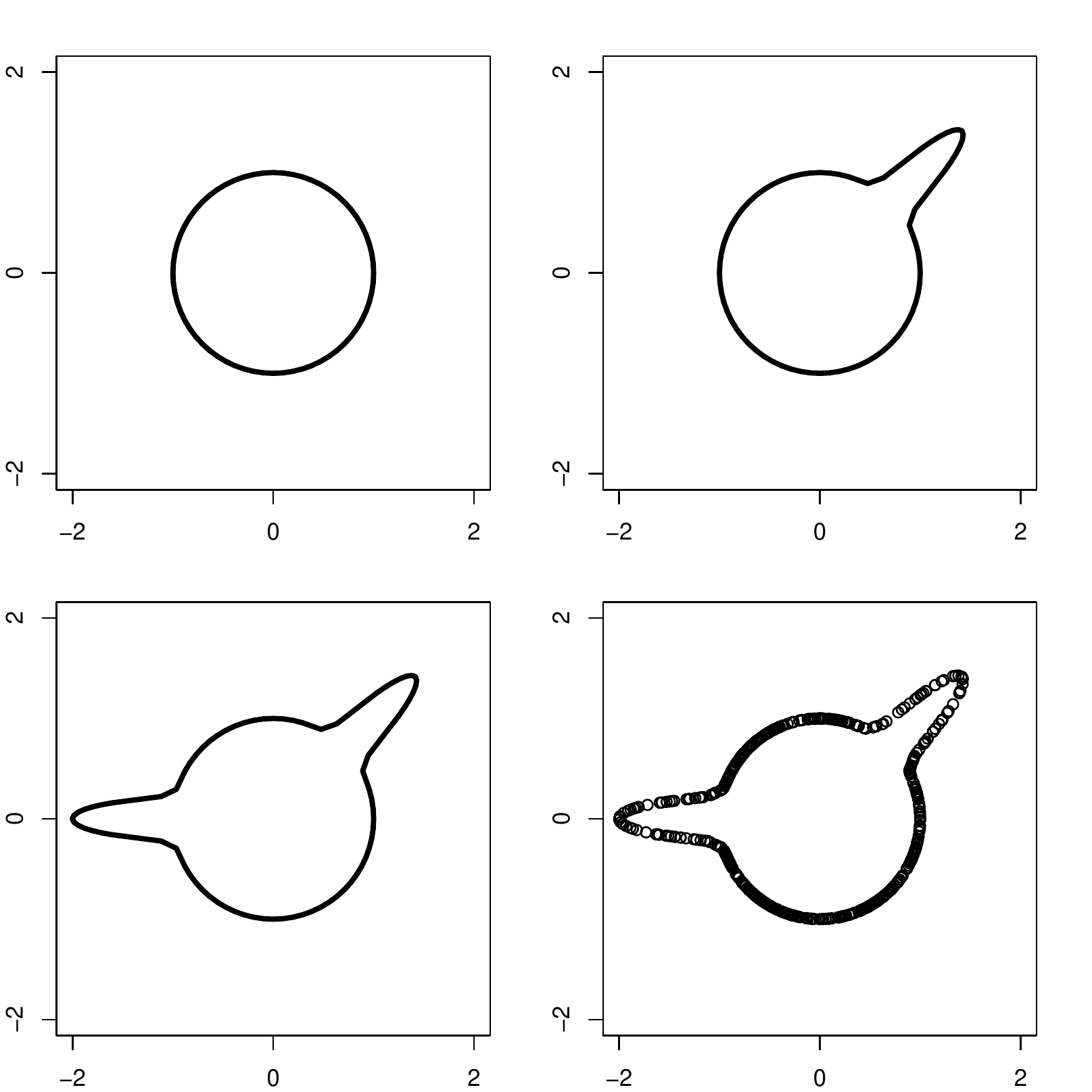}}
\caption{Constructing a 2-dimensional contour. The top left plot shows a base of type 1, a circle of radius 1.
The top right shows the base with one Gaussian bump of type 3 in direction $(\sqrt{2}/2,\sqrt{2}/2)$, the bottom left shows
the final contour with another Gaussian bump in direction (-1,0).  The bottom right plot shows
a sample of size $n=1000$ from this contour using the  method described in Section~\ref{sec:tess.integ.sim}.  }\label{fig:construct2d}
\end{figure}

% One possible approach to define more general functions on spheres
%is to use splines on spheres as described by
%\cite{ferreira:steel:2005}, see \cite{wahba:1981} and \cite{taijeron:gibson:chandler:1994}.
Sums of the first three types allow us to describe star shaped contours, see Figure~\ref{fig:construct2d}.
Inverses of sums of the last two types allow us to consider contours that are familiar unit balls,
or generalized unit balls, or sums of such shapes.
An implementation of this construction is given in a new R package \Rpkg{gensphere}.  The
R statements used in this example are given in the Appendix.

\begin{figure}
\myfig{[width=\textwidth]{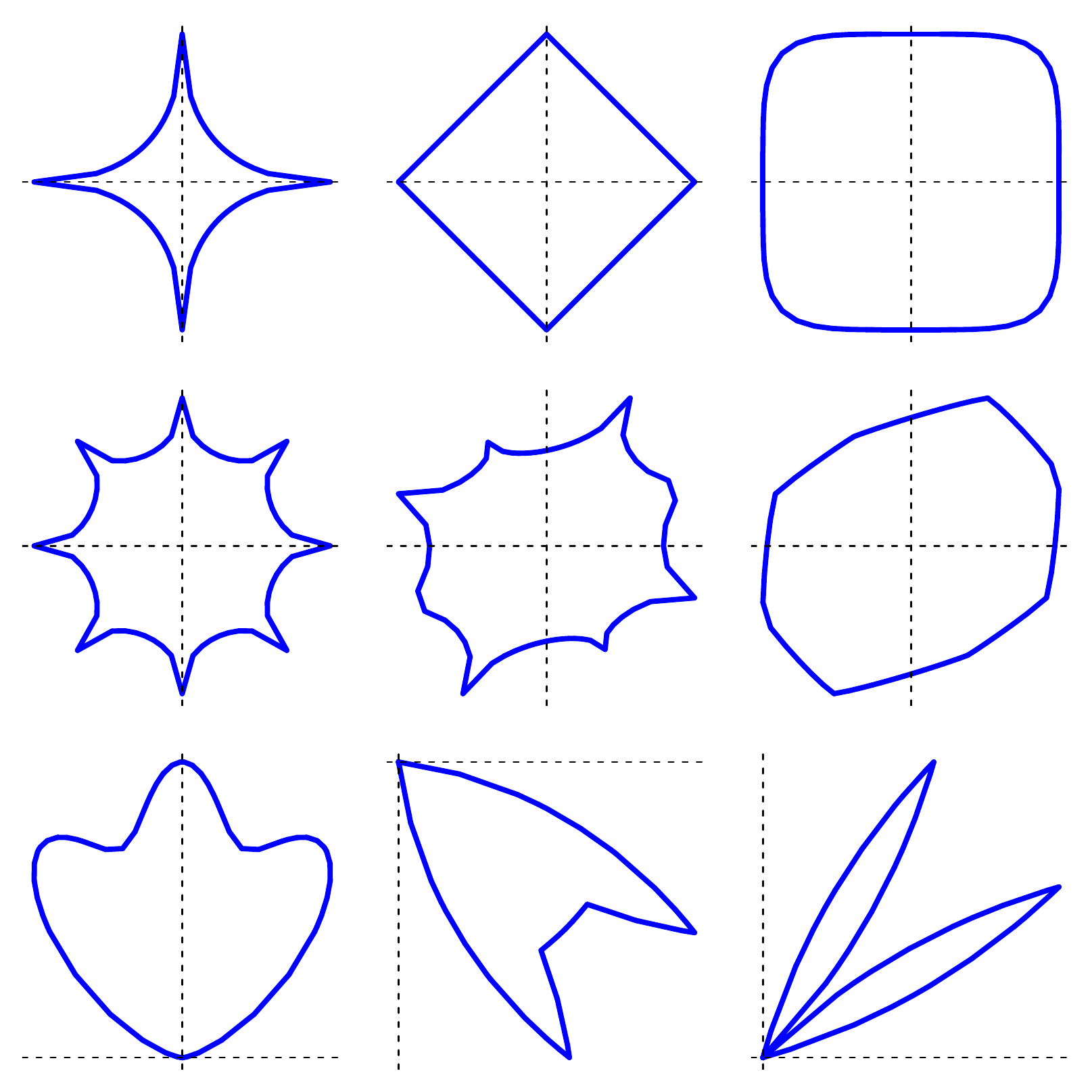}}
\caption{A selection of contours made from the different types of terms.  See the text for
a description. }\label{fig:misc2dcontours}
\end{figure}

\begin{figure}
\myfig{[height=6in]{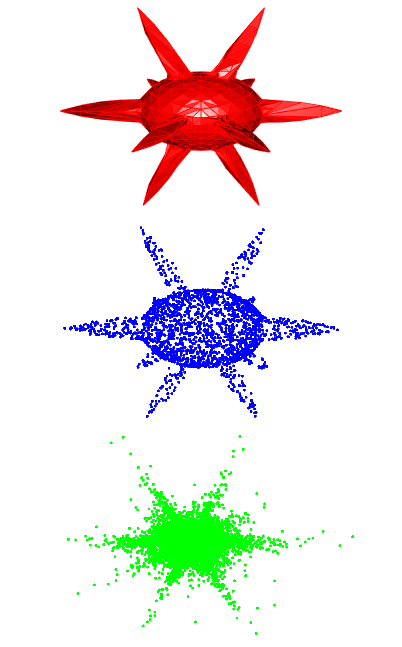}}
\caption{A 3D star-shaped region with one term of type 6 and 10 terms of type 2.
The top plot shows the contour, the middle shows a sample of size 2500 from the contour,
the bottom shows a sample of size 10000 from the generalized spherical distribution
given by this contour and  a $\Gamma(3,1)$ radial term $R$.}\label{fig:star3d}
\end{figure}

It is relatively easy to add new types of terms to this list if other contours are of interest.
However this set of basic shapes can model a wide range of shapes, including contours
supported on a cone.  Figure~\ref{fig:misc2dcontours} shows nine examples.  The top row
shows $\ell_p$ balls with $p=1/2$, $p=1$, and $p=5$.  The middle row starts with a
contour made up of an $\ell_p$ ball with a $p=0.3$ and a copy of that rotated by $\pi/4$,
the rotation done by using a generalized $\ell_p$ norm with $A$ a rotation matrix.
The next two plots show generalized $\ell_p$ balls with $A=(1,1; 1,-4; 1,3; 5,-3)$ and
$p=1/2$ (middle) and $p=1.1$ (right).  The last row shows contours supported on a cone.
The left plot is the sum of three Gaussian bumps of type 3, each centered at $(\cos \theta, \sin \theta)$,
$\theta=\pi/4,\pi/2,3 \pi/2$ and $\sigma=0.3$.  The middle plot has two type 2 cones,
at angles $-\pi/6$ and $-\pi/3$ with $\sigma=0.4$.  The last graph  also has two cones,
centered at $\pi/6$ and $\pi/3$, with $\sigma=0.25$.  Any of the contours that have
a corner or cusp on a ray will generate a density surface
with a ridge along that ray.
A more complicated three dimensional example with 11 terms in the definition
of $c(\cdot)$ is given in Figure~\ref{fig:star3d}.

\begin{figure}
\myfig{[height=3in]{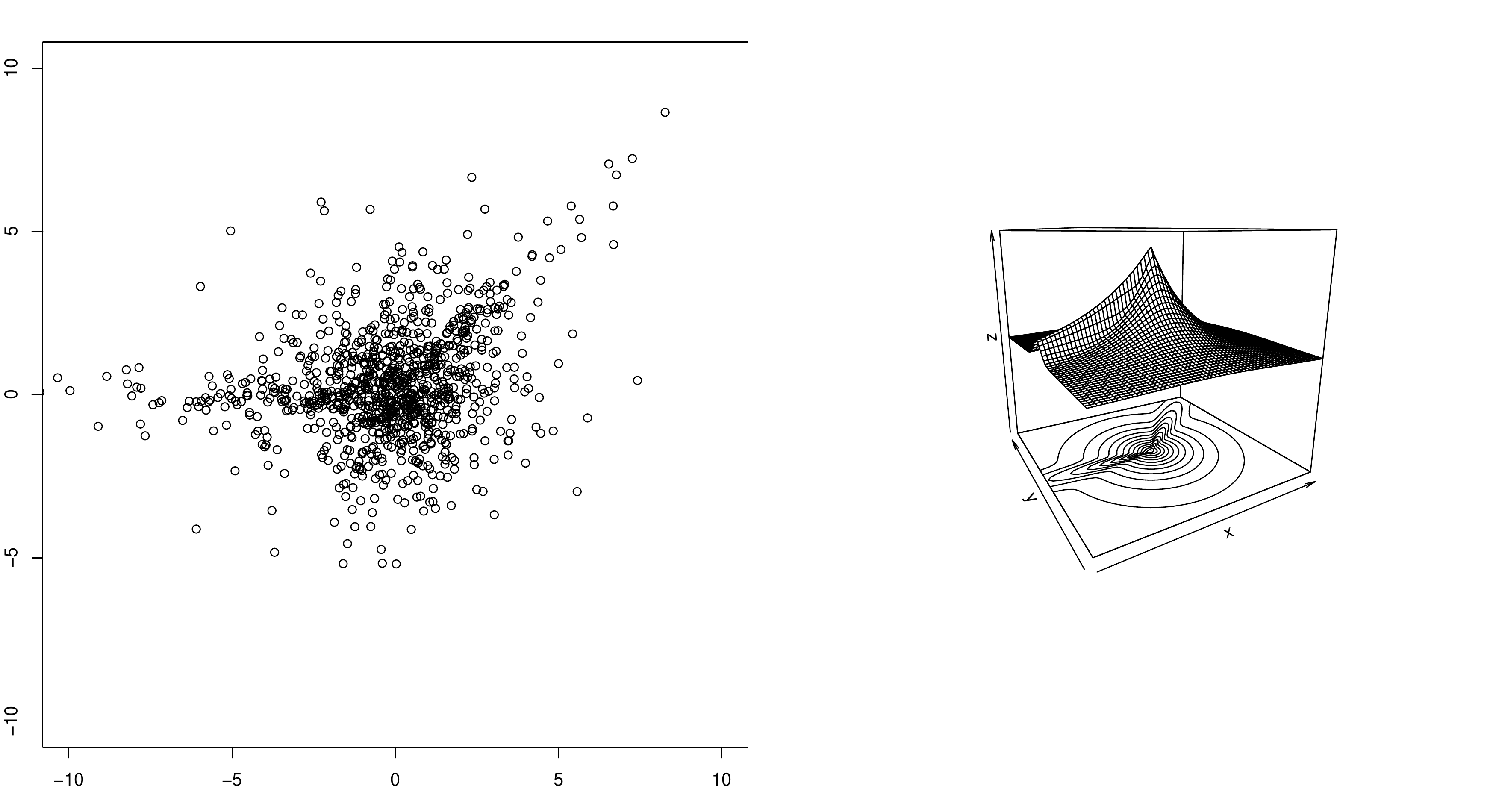}}
\caption{On the left is a sample from the generalized spherical distribution based on the
contour in Figure~\ref{fig:construct2d} and a $\Gamma$(2,1) radial term.
On the right, is a surface plot of the density $f(x,y)$.  }\label{fig:samp.contour2d}
\end{figure}

%-----------------------------------------------------------------------------
\subsection{Choice of $R$}

In general, $g(r)$ can be any nonnegative integrable function.
In most applications one wants $0 < g(0) < \infty$ and $g(r)$ decreasing for $r > 0$.
If $g(0)=0$, the density surface given by (\ref{eq:f.def}) will have a ``well''
at the origin; if $g(0)=+\infty$, then the density blows up at the origin.
If $g(\cdot)$ oscillates, then the density surface will have radial ``waves''
emanating out from the origin.  If $R$ has bounded support, then $\boldX$ will
have bounded support.  The radial decay of $R$ determines the decay of $f(\cdot)$ on $\reals^d$.

The gamma distribution give a family of distributions that
can by used to get generalized spherical distributions with light tails.
If a $\Gamma(d,1)$ law is used for $R$, then $h(r)=\Gamma(d)^{-1} r^{d-1} \exp(-r)$, so
$g(r)=k_{\mathcal{C}} r^{1-d} h(r) = (k_{\mathcal{C}}/\Gamma(d)) \exp(-r)$,
which  is finite at the origin and monotonically decreasing.
If one wants heavy tails for $\boldX$, then some possibilities for $R$
are Fr\'echet, Pareto and multivariate stable amplitude.  (The latter is
defined in \cite{nolan:2013} by $R=|\boldZ|$, where
$\boldZ$ is radially symmetric/isotropic $\alpha$-stable in $d$-dimensions.
Numerical methods to calculate the density $h(r)$ of $R$ and
simple ways to simulate are given in the reference.)

\begin{figure}
\myfig{[height=5in]{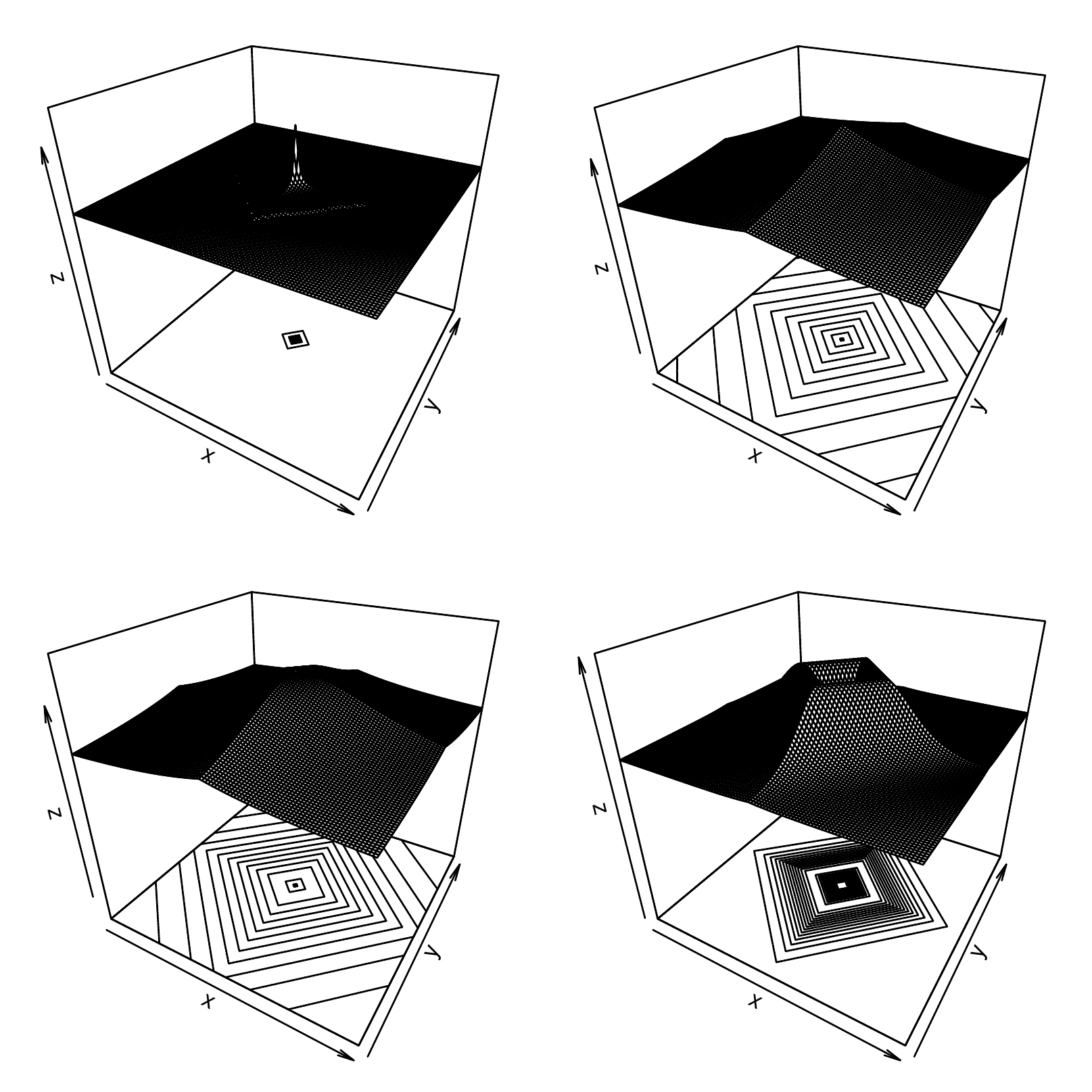}}
\caption{Density surface for generalized spherical distributions
with the same diamond shape and different radial term $R$
as specified in the text.}\label{fig:r.demo}
\end{figure}

Figure~\ref{fig:r.demo} shows the effect that the choice of $R$ has.
In all cases, the base contour is the unit ball in $\ell_1$, a diamond shape.  At the upper left,
$R$ is a uniform r.v. on (0,1).  In this case, $g(0)=+\infty$ and the density
has a spike at the origin and bounded support on the diamond. At the top right,
$R \sim \Gamma(2,1)$, so $g(0)=1$ and the distribution has unbounded support
with light tails.  At the lower left, $R$ is the $\alpha=1$ stable amplitude in
$d=2$ dimensions; here $g(0)$ is finite and the distribution has heavy tails.
The bottom right plot is with $R \sim \Gamma(5,1)$, so $g(0)=0$ and the distribution
has a well at the origin and unbounded support with light tails.

%-----------------------------------------------------------------------------
\section{Contours: tessellating, integrating and simulating}\label{sec:tess.integ.sim}

A large part of the technical complexity of working with generalized spherical laws
is in evaluating the norming constant $k_{\mathcal{C}}$ in (\ref{eq:norming.constant})
and simulating from the contour $\mathcal{C}$.
The \Rpkg{gensphere} package uses two other recent R packages for these problems:
\Rpkg{SphericalCubature} \cite{SphericalCubature:2015} and \Rpkg{mvmesh} \cite{mvmesh:2015}.

\Rpkg{SphericalCubature} numerically integrates a function on a $d$-dimensional sphere.
Given a tessellation of the sphere in $\reals^d$, it uses adaptive integration to integrate
over the $(d-1)$-dimensional surface to evaluate $k_{\mathcal{C}}$.
If the integrand function is
smooth and the tessellation is reasonable, then the numerical integration is
accurate in modest dimensions, say $d=2,3,4,5,6$.  However, when the
integrand function has abrupt changes, numerical techniques can miss
parts of the integral.  This is even a problem in dimension
2, where the integration is a one dimensional problem.  One way to
deal with this is to work with tessellations that focus on the
places where the integrand is not smooth.  In complete
generality, this is hard to do.  However, in evaluating
integral (\ref{eq:norming.constant}) for one of the contours
described above, we have an implicit description of where
the contour changes abruptly.

The \Rpkg{mvmesh} package is used to define multivariate meshes, e.g. a collection of
vertices and grouping information that specify a list of simplices
that approximate a contour.  The first place where \Rpkg{mvmesh} is used in \Rpkg{gensphere} is
to give a grid on the sphere $\sphere$ in $d$-dimensions,
e.g. the top left plot in Figure~\ref{fig:construct2d}.  \Rpkg{mvmesh} has a
function UnitSphere that computes an approximately equal surface area approximation
to a hypersphere in dimension $d$.  It takes a parameter $k$ to say how many recursive subdivisions
are used in each octant.  Then this tessellation is refined by adding points on the to
the sphere centered on the places where the contour has bumps, e.g. the
cone and Gaussian bumps (type 2 and 3).  Then the new points are
combined with the original tessellation of the sphere to get a refined
tessellation of the sphere that includes these key points.

It is at this point that the \Rpkg{SphericalCubature} package is used to
evaluate the integral (\ref{eq:g.integ}).  In addition to the
estimate of the integral, we use an option in the
adaptive integration routine to return the partition used in the
multivariate cubature, along with the estimated integral over each
simplex. The reasoning is that the integration routine is subdividing
regions  where the integrand is changing quickly to get a better
estimate of the integrand. This subdivision should make the tessellation
more closely approximate the contour.  We now have the final tessellation
of the unit sphere, an estimate of the integral (\ref{eq:norming.constant})
over each of the simplices, and an estimate of the norming constant, e.g.
sum of these just mentioned values.

Now the tessellation of the contour is defined by deforming the tessellation
on the sphere to the contour: each partition point $\bolds \in \sphere$
gets mapped to $c(\bolds) \bolds$ on the contour.  The grouping information
from the spherical tessellation is inherited by the contour tessellation.
This tessellation is returned as an S3 object of class ``mvmesh''.
This object contains the vertices, the grouping information, and a list of
all the simplices $S_1,S_2,\ldots,S_k$ in the tessellation.  One advantage of this is that
the plot method the \Rpkg{mvmesh} package can plot the contours in 2 and 3 dimensions.
This process of refining the tessellation has two purposes: (a) get a more accurate
estimate of the norming constant by focusing the numerical integration routine
on regions where the integrand changes rapidly and (b) get a more accurate
tessellation of the contour. Each step of this process can add more simplices,
with the goal of capturing key features of the contour.  For example, the
contour in Figure~\ref{fig:star3d} started with 512 simplices in the tessellation
of the sphere in $\reals^3$ with $k=3$, adding the points on the cone brought the number up to 888
simplices, and after the adaptive cubature routine subdivision there were 2284
simplices.

\bigskip
Exact simulation from a surface is a challenging problem and general methods
%There are some techniques to sample from the manifold, e.g.  \cite{diaconis:holmes:shahashahani:2012}
%and  \cite{saucan:appleboim:zeevi:2007}.  These
are difficult to apply for complicated contours like our star shaped regions.
We now describe an approximate method based on the above tessellation.
Recall that the above process gives us a list of simplices $S_1,\ldots,S_k$ and
associated weights $w_1,\ldots,w_k$ that are estimates of the integral
on each simplex; $w_j$ is an estimate of the surface area of the contour
approximated by simplex $S_j$.

The simulation routine is now straightforward:
\begin{enumerate}
  \item select an index $j$ with probability proportional to $w_j$.
  \item simulate a point $\boldu$ that is uniformly distributed on the unit simplex in $d$-dimensions.
     This is standard: simulate a Dirichlet distribution with parameter $\boldalpha=(1,1,\ldots,1)$,
     e.g. let $E_1,\ldots,E_d$ i.i.d. standard exponential random variates and
      set $\boldu=(E_1,\ldots,E_d)/ \left( \sum_{i=1}^d E_i \right)$.
  \item map the point $\boldu$ to the simplex $S_j$ using the coordinates of
  $\boldu$ as barycentric coordinates: $\boldZ = \boldu^T S_j$.
  \item simulate $R$ from the radial distribution with density $h(r)$.
  \item return the value $\boldX = R \boldZ$.
\end{enumerate}

\begin{figure}
\begin{center}
\myfig{[width=1.1\textwidth]{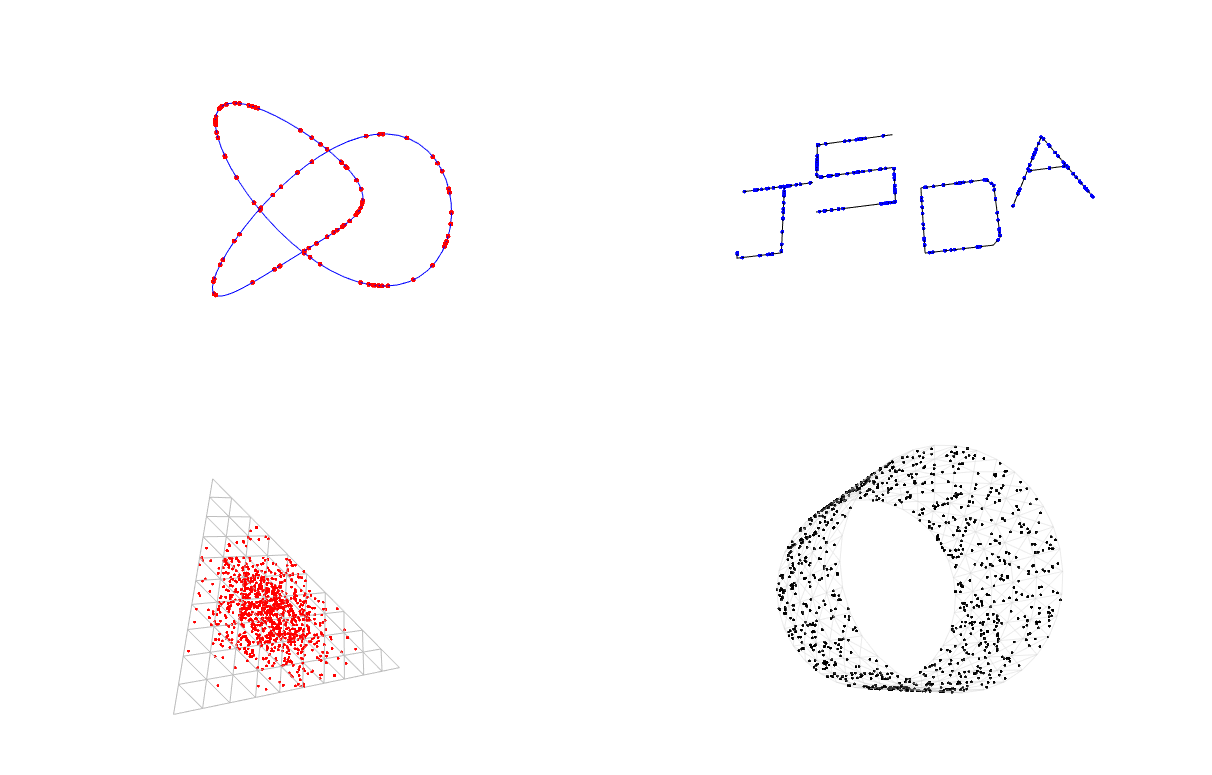}}
\end{center}
\caption{Simulation from general sets.
Top left is a trefoil knot, the points are sampled uniformly.  Top right has the
letters JSDA embedded in $\reals^3$ and points are sampled uniformly on them.
Bottom left has points sampled on the unit simplex according to a density
$\exp(-c |\boldx-(\frac 1 3, \frac 1 3,\frac 1 3) |^2 )$.  The last plot shows points sampled
uniformly from a hollow tube.}\label{fig:rtessellations}
\end{figure}

This method works in any dimension and the first three steps is adaptable to a wide variety of shapes,
more than just the contours described above. For the generalized spherical distributions the weights are proportional
to surface area, e.g. the bottom right
plot in Figure~\ref{fig:construct2d}, but they can be assigned in any way.
Figure~\ref{fig:rtessellations} illustrates
some examples with different shapes and where the weights are defined in different ways.
In all cases the points $\boldZ$ are sampled from the simplex faces; too work
well the tessellation should closely model the shape of the surface of interest.

%-----------------------------------------------------------------------------
%\section{Concluding remarks}

\bigskip
The subdivision process, including the numerical cubature is the
slowest part of the process. This is done in the R function cfunc.finish,
which finishes the definition of a contour by performing these calculations
and saving the results in an object of class ``contour.function''.
For the example in Figure~\ref{fig:star3d}, it can take several
minutes to complete the construction.  The simulation and density calculations
are fast, taking only a few seconds.  Producing the graphs can be slow; the top
plot in Figure~\ref{fig:star3d} with 2284 simplices took over 1/2 an hour to plot.

In contrast, once the tessellation is produced, density calculations and simulations
are quite fast: to evaluate a density at 10000 points takes less than a second\footnote{Times are for an
Intel i5-4460 CPU at 3.20 GHz}, and to simulate $n=10000$ random vectors takes less than a second
in a bivariate example with two terms.

In principle, the methods described here work in any dimension; in practice
the numerical challenges, particularly evaluating the integral in (\ref{eq:norming.constant})
and the time needed to work limit us as the dimension increases. 

%Higher dimensional examples...

% Separate into two parts: (1) determining the radial $g(\cdot)$ and (2) determining the contour $\mathcal{C}$.

% Pick $g(\cdot)$.

% Pick a class of models, e.g. the number and types of terms allowable in (\ref{eq:f.def}).
% Then one can use (\ref{eq:f.def}) to compute the density $f(\boldx)$ and therefore
% the likelihood.  One can numerically maximize the likelihood.  For star shaped regions,
% can allow one term of type 1 (to guarantee $c(\cdot) > 0$)
% and $m$ terms of type 2 to determine the best location and scale parameters.
% Can vary $m$ and use AIC to select optimal number of cones.

% --------------------------------------------------------------------------
\appendix
\section{Appendix}

Here are the R statements used
to produce Figures~\ref{fig:construct2d} and \ref{fig:samp.contour2d} from the R package \Rpkg{gensphere}.

\begin{verbatim}
# define a new contour function (cfunc)
cfunc <- cfunc.new(d=2)
cfunc <- cfunc.add.term( cfunc,"constant",k=1)
cfunc <- cfunc.add.term( cfunc,"proj.normal",
             k=c( 1, sqrt(2)/2, sqrt(2)/2, 0.1) )
cfunc <- cfunc.add.term( cfunc,"proj.normal",
             k=c( 1, -1,0, 0.1) )
cfunc <- cfunc.finish( cfunc, k=4 )

# define a generalized spherical distribution with the
# above contour and a gamma radial component
rradial <- function( n ) { rgamma( x, shape=2 ) }
dradial <- function( x ) { dgamma( x, shape=2 ) }
dist <- gensphere( cfunc, dradial, rradial, g0=1 )

# simulate from the generalized spherical distribution
x <- rgensphere( 1000, dist )

# compute the density
xx <- matrix(1:8,nrow=2)
dgensphere( xx, dist )

# simulate from the tessellation of the contour
x <- rtessellation( n=500, (cfunc$tessellation)$S,
     cfunc$tessellation.weights )
\end{verbatim}

\section{Competing Interests}

I confirm that I have read SpringerOpen's guidance on competing interests and  have no competing interests in the manuscript.

% --------------------------------------------------------------------------

\bibliographystyle{chicago}
\bibliography{GeneralizedSpherical}

\begin{thebibliography}{}

\bibitem[\protect\citeauthoryear{Arnold, Castillo, and Sarabia}{Arnold
  et~al.}{2008}]{arnold:castillo:sarabia:2008}
Arnold, B.~C., E.~Castillo, and J.~M. Sarabia (2008).
\newblock Multivariate distributions defined in terms of contours.
\newblock {\em Journal of Statistical Planning and Inference\/}~{\em
  138\/}(12), 4158 -- 4171.

\bibitem[\protect\citeauthoryear{Balkema and Nolde}{Balkema and
  Nolde}{2010}]{balkema:nolde:2010}
Balkema, G. and N.~Nolde (2010).
\newblock Asymptotic independence for unimodal densities.
\newblock {\em Adv. in Appl. Probab.\/}~{\em 42\/}(2), 411--432.

\bibitem[\protect\citeauthoryear{Fern{\'a}ndez, Osiewalski, and
  Steel}{Fern{\'a}ndez et~al.}{1995}]{fernandez:osiewalski:steel:1995}
Fern{\'a}ndez, C., J.~Osiewalski, and M.~F.~J. Steel (1995).
\newblock Modeling and inference with $v$-spherical distributions.
\newblock {\em Journal of the American Statistical Association\/}~{\em
  90\/}(432), 1331--1340.

\bibitem[\protect\citeauthoryear{Kamiya, Takemura, and Kuriki}{Kamiya
  et~al.}{2008}]{kamiya:takemura:kuriki:2008}
Kamiya, H., A.~Takemura, and S.~Kuriki (2008).
\newblock Star-shaped distributions and their generalizations.
\newblock {\em J. Statist. Plann. Inference\/}~{\em 138\/}(11), 3429--3447.

\bibitem[\protect\citeauthoryear{Nolan}{Nolan}{2013}]{nolan:2013}
Nolan, J.~P. (2013).
\newblock Multivariate elliptically contoured stable distributions: theory and
  estimation.
\newblock {\em Computational Statistics\/}~{\em 28\/}(5), 2067--2089.

\bibitem[\protect\citeauthoryear{Nolan}{Nolan}{2015a}]{mvmesh:2015}
Nolan, J.~P. (2015a).
\newblock {\em mvmesh: Multivariate Meshes and Histograms in Arbitrary
  Dimensions}.
\newblock R package version 1.1, on CRAN.

\bibitem[\protect\citeauthoryear{Nolan}{Nolan}{2015b}]{SphericalCubature:2015}
Nolan, J.~P. (2015b).
\newblock {\em SphericalCubature: Numerical Integration over Spheres and Balls
  in n-Dimensions; Multivariate Polar Coordinates}.
\newblock R package version 1.1, on CRAN.

\bibitem[\protect\citeauthoryear{Rattihalli and Basugade}{Rattihalli and
  Basugade}{2009}]{rattihalli:basugade:2009}
Rattihalli, R.~N. and A.~B. Basugade (2009).
\newblock Generation of densities using contour transformations.
\newblock {\em J. Indian Statist. Assoc.\/}~{\em 47\/}(1), 63--90.

\bibitem[\protect\citeauthoryear{Rattihalli and Patil}{Rattihalli and
  Patil}{2010}]{rattihalli:patil:2010}
Rattihalli, R.~N. and P.~Y. Patil (2010).
\newblock Generalized {$v$}-spherical densities.
\newblock {\em Comm. Statist. Theory Methods\/}~{\em 39\/}(19), 3568--3583.

\end{thebibliography}

\end{document}